# Superconductivity in diamond.

E. A. Ekimov[1], V. A. Sidorov[1,2], E. D. Bauer[2], N. N. Mel'nik[3], N. J. Curro[2], J. D. Thompson[2], and S. M. Stishov[1]

[1]*Vereshchagin Institute for High Pressure Physics, Russian Academy of Sciences, 142190 Troitsk, Moscow region, Russia*

[2]*Los Alamos National Laboratory, Los Alamos, New Mexico 87545, USA*

[3]*Lebedev Physics Institute, Russian Academy of Sciences, 117924 Moscow, Russia*

**Diamond is an electrical insulator well known for its exceptional hardness. It also conducts heat even more effectively than copper, and can withstand very high electric fields[1]. With these physical properties, diamond is attractive for electronic applications[2], particularly when charge carriers are introduced (by chemical doping) into the system. Boron has one less electron than carbon and, because of its small atomic radius, boron is relatively easily incorporated into diamond[3]; as boron acts as a charge acceptor, the resulting diamond is effectively hole-doped. Here we report the discovery of superconductivity in boron-doped diamond synthesized at high pressure (8-9 GPa) and temperature (2,500–2,800 K). Electrical resistivity, magnetic susceptibility, specific heat and field-dependent resistance measurements show that boron-doped diamond is a bulk, type-II superconductor below the superconducting transition temperature $T_c \approx 4$ K; superconductivity survives in a magnetic field up to $H_{c2}(0) \geq 3.5$ T. The discovery of superconductivity in diamond-structured carbon suggests that Si and Ge, which also form in the diamond structure, may similarly exhibit superconductivity under the appropriate conditions.**

With their potential for electronic applications as microchip substrates, high efficiency electron emitters, photodetectors and transistors, diamond and carrier-doped diamond have been studied extensively[3–6]. The extremely short covalent bonds of carbon atoms in diamond give diamond many of its desirable properties, but also constrain geometrically which dopants can be incorporated and their concentration. Because of its small atomic radius compared to other potential dopants, boron is readily incorporated into the dense ($1.763 \times 10^{23}$ atoms cm$^{-3}$) diamond lattice. Boron dopes holes into a shallow acceptor level close to the top of the valence band that is separated from the conduction band of diamond by $E_g \approx 5.5$ eV. Electrical transport studies of B-doped diamond, including high-pressure synthesized crystals and CVD (chemical vapour deposition) films, find that low boron concentrations $n \approx 10^{17}$–$10^{19}$ cm$^{-3}$ give a semiconducting conductivity with an activation energy of ~0.35 eV (refs 7–11). Increasing the concentration to $10^{20}$ cm$^{-3}$ gradually decreases the activation energy[9,10], and for $n \geq 10^{20}$ cm$^{-3}$, the electrical conductivity acquires metallic-like behaviour near room temperature[8–11] that signals an insulator–metal transition near this concentration. A metallic-like conductivity has not been found, however, at low temperatures for any presently available B concentration, which has reached $(2–3) \times 10^{21}$ cm$^{-3}$ (refs 8–11).

We have studied B-doped diamond synthesized by reacting B$_4$C and graphitic carbon at pressure, 8–9 GPa, and temperature, 2,500–2,800 K, for ~5 s. Under these conditions, polycrystalline diamond aggregates 1–2 mm in size formed at the interface between graphite and B$_4$C. All the diamond aggregates had a metal-like lustre. Scanning electron microscopy (SEM) showed that the diamond grain size was a few micrometres (Fig. 1). The samples were characterized by X-ray diffraction, micro-Raman spectroscopy, electrical resistivity, magnetic susceptibility and calorimetry measurements. Consistent with an expanded lattice deduced from the X-ray spectrum given in Fig. 2a, Raman spectra of our samples (Fig. 2b) are very similar to those observed for the most heavily boron-doped CVD diamond films[11]. (From the weak zone-centre phonon line at ~1,300 cm$^{-1}$ and the general form of the spectra[11], our samples should have a carrier concentration $n \geq 2 \times 10^{21}$ cm$^{-3}$. For normal diamond, a zone-centre phonon line is observed at 1,332 cm$^{-3}$.) Quantitative estimates of the boron content in our samples were made using NMR and inductively coupled mass spectrometry. Within experimental uncertainty, both methods yielded the same value of 2.8±0.5% for the total B content. Inductively coupled mass spectrometry on a second sample also gave the same B content. Good agreement between these two techniques, one of which is volume sensitive (NMR) and the other of which probes only a surface layer of ~1,000 Å, suggests that B atoms are distributed rather uniformly in the sample. Neither technique, however, is sensitive to the local environment of the B atoms. Assuming that all the B atoms are incorporated into the diamond lattice, the upper limit of the charge carrier concentration is $4.9 \times 10^{21}$ cm$^{-3}$, a value somewhat larger than the maximum previously reported. This estimate decreases to $4.6 \times 10^{21}$ cm$^{-3}$, assuming the sample contains 2% B$_4$C.

The temperature ($T$)-dependent resistivity ($\rho$) of one of these B-doped samples is plotted in Fig. 3a. Measurements of current-voltage characteristics at various temperatures confirmed an ohmic response. In the range 230–300 K, $\partial \rho / \partial T > 0$ as expected for a metal, but below 230 K the resistivity increases weakly with decreasing temperature. Near 4 K, $\rho(T)$ starts to fall rapidly and reaches an immeasurably small value below 2.3 K as shown in the inset of Fig. 3a. The resistive variation below 4 K is typical of an inhomogeneous superconductor, in this case with the inhomogeneity probably arising from a non-uniform distribution of boron in the diamond lattice. As also indicated in the inset of Fig. 3a, the application of hydrostatic pressure produces a linear decrease in the resistive mid-point transition temperature $T_c$ without any additional significant broadening of the transition. The relatively slow decrease of $T_c$ with pressure $P$

($\partial T_c/\partial P$=−6.42×10$^{-2}$ K GPa$^{-1}$, $\partial \ln T_c/\partial P$=−2.79×10$^{-2}$ GPa$^{-1}$) contrasts with positive $\partial T_c/\partial P$=0.05 K GPa$^{-1}$ reported[12] for elemental boron for $P$>180 GPa, ruling out the possibility that the zero-resistance state in B-doped diamond is due to free boron. As expected for an inhomogeneous superconductor, applying a magnetic field slightly broadens the transition and shifts it to lower temperatures (Fig. 3b). The resistive mid-point $T_c$ decreases at a rate $\partial H_{c2}/\partial T_c$=−1.7 T K$^{-1}$, from which we estimate the $T$=0 upper critical field $H_{c2}(0)$=3.4 T using the standard relationship[13] for a dirty type-II superconductor $H_{c2}(0)$=−0.69 (d$H_{c2}$/d$T|_{Tc}$)$T_c$. As evident in the inset of Fig. 3b, this estimate is a lower limit because the onset of the resistive transition is still near 1.7 K in a field of 4 T. With $H_{c2}(0)$=3.4 T and the relation $\xi_{GL}$=[$\Phi_0/2\pi H_{c2}(0)$]$^{1/2}$, where $\Phi_0$ = hc/2e is the quantum of magnetic flux, the Ginzburg-Landau coherence length $\xi_{GL}$=100 Å.

Resistance measurements are unable to determine if superconductivity arises from the crystal bulk, surface, or filaments of zero-resistance material; however, magnetic susceptibility measurements allow more definitive conclusions. As shown in Fig. 4, there is a strong diamagnetic response in the a.c. magnetic susceptibility of these boron-doped diamond samples below 2.3 K where the resistance drops to zero. At 1.1 K, the diamagnetic response for various B-doped samples ranged from 25% to 60% of the value measured for a superconducting indium sample in the same coil system. An absolute value of magnetic susceptibility for two separate samples, measured by SQUID (superconducting quantum interference device) magnetometry, agrees with the above estimates (Fig. 4), and, further, finds a diamagnetic response on cooling in a d.c. field of 5 Oe. These results indicate that the superconductivity is not filamentary. Moreover, magnetization measurements versus field (inset of Fig. 4) find the magnetic hysteresis expected in a bulk superconductor with flux pinning.

A specific heat anomaly at $T_c$ would provide the most definitive evidence that superconductivity develops in the bulk of B-doped diamond. Assuming that the effective mass of charge carriers is the free electron mass $m_e$, we estimate the electronic contribution to the specific heat as C = $\gamma$T. $\gamma$=7.6×10$^{-4}$ J mol$^{-1}$ K$^{-2}$ from the free electron relationships $k_F$=(3$\pi^2 n$)$^{1/3}$ and $\gamma$=$\pi^2 n k_B^2 m_e/\hbar^2 k_F^2$, where $k_F$ is the Fermi momentum, $n$ is the carrier density, and $k_B$ is Boltzmann's constant. For a weak coupled BCS superconductor, the jump in specific heat C at $T_c$ is $\Delta C/\gamma T_c$=1.43, which implies $\Delta C$=2.5×10$^{-3}$ J mol$^{-1}$ K$^{-1}$ and for our largest 2-mg sample, then, $\Delta C$=4.2×10$^{-7}$ J K$^{-1}$. This is a very small value, essentially equal to the noise level of a commercial Quantum Design Physical Property Measurement System. We have measured the specific heat of this 2-mg sample in such a device and find a barely discernable anomaly near 2.3 K.

For phonon-mediated superconductivity, $T_c$ is given approximately by $T_c$=($\theta_D$/1.45)exp(−1.04(1+$\lambda$)/$\lambda$), where $\theta_D$ is the Debye temperature and $\lambda$ is a measure of electron–phonon coupling[14]. The high Debye temperature of diamond ($\theta_D$=1,860 K) suggests that carrier doping the diamond lattice could produce a relatively high $T_c$. Using this value of $\theta_D$ and $T_c$=2.3 K, we estimate $\lambda$=0.2, indicative of weak electron–phonon coupling. This somewhat small value of $\lambda$ can be understood qualitatively from the relation $\lambda=N(0)<I^2>/(M<\omega^2>)$, where $N(0)$ is the density of electronic states at the Fermi energy, $<I^2>$ is the average square of the electron–phonon matrix element, $M$ is the ionic mass and $<\omega^2>$ is a characteristic phonon frequency averaged over the phonon spectrum[14]. Although the ionic mass is light, the strong covalent bonds in diamond and low concentration of charge carriers imply a small $N(0)$ and large $<\omega^2>$.

Superconductivity in doped semiconductors is an intriguing issue for theory[15–19], but experimentally only a few superconducting semiconductors have been discovered. Bulk superconductivity with $T_c \approx$0.1–0.5 K was found[20] in self-doped GeTe and SnTe at carrier densities $n \approx 10^{21}$ cm$^{-3}$ and in doped SrTiO$_3$ ($T_c \approx$0.05–0.25 K at $n \approx 10^{19}$–$10^{21}$ cm$^{-3}$). There have been no reports until now of superconductivity in group-IV semiconductors with the diamond structure—silicon, germanium and their alloys—though it was predicted[16] to exist at very low temperatures. The discovery of bulk superconductivity in B-doped diamond should stimulate a fresh look at this old problem, which is posed more straightforwardly in B-doped diamond than in the more complex Si- or Ge-based clathrates, which are the first superconductors[21,22] built from a network of tetrahedral covalent bonds having bond lengths comparable to those of cubic diamond. Few, if any, measurements of B-doped diamond have been made below 4 K (ref. 8), so that superconductivity may have been missed in samples prepared by others with lower B concentrations. A systematic study seems worthwhile. Our results also suggest that homogeneously doping diamond with B concentrations of (4–5)×$10^{21}$ cm$^{-3}$ should produce sharp transitions near 4 K and upper critical fields above 4 T.

**Methods**

**Sample preparation and analysis**

Starting materials, graphite discs containing ≤ 500 p.p.m. impurities and B$_4$C powder containing less than 3,000 p.p.m. impurities (mainly Fe≤2,000 p.p.m., and Si≤1,000 p.p.m.), were placed in a graphite heater assembly. Impurity content in the starting materials was determined by X-ray fluorescence and spectroscopic methods, both showing identical results. Starting materials were annealed in vacuum to remove moisture. The heater was insulated by boron nitride sleeves from the gasket material (lithographic stone, mainly CaCO$_3$+SiO$_2$) and opposed tungsten carbide anvils (with 6% Co) of a high-pressure toroid apparatus (ref. 23). Molybdenum disks separated cylindrical parts of the graphite heater and anvils. Impurity content in the synthesized diamond was determined by X-ray fluorescence, with special attention to Ca, Co, Fe and Si, which could arise from the gasket, anvils and B$_4$C. Total content of these elements does not exceed 500 p.p.m. The boron content in our diamond samples was estimated in two ways. NMR spectra were measured at room temperature and peaks corresponding to $^{13}$C and $^{11}$B isotopes were observed. The ratio of boron/carbon atoms in the sample was calculated on the basis of the peak areas and the known natural abundance of $^{11}$B/$^{13}$B and $^{13}$C/$^{12}$C isotopes. The second method used a high-resolution inductively coupled mass spectrometer (ThermoFinnigan-MAT 'Element 2') equipped with an ultraviolet laser ablation microprobe. The samples also were characterized by X-ray diffraction and micro-Raman spectroscopy. X-ray diffraction studies were made with a BRUKER-AXS diffractometer and monochromatized Mo radiation. Micro-Raman spectroscopy was performed using a U-1000 spectrometer. The laser exiting radiation (488 nm) was focused to a spot ~20 μm in diameter. The Raman spectra were recorded in backscattering geometry with a resolution better than 1 cm$^{-1}$.

**Electric and magnetic measurements**

Resistivity measurements were made in a standard four-terminal configuration using an LR700 a.c. resistance bridge. Silver epoxy attached 25-μm Pt wires to the samples. Measurements in the range 1.1–300 K were performed in a $^4$He cryostat. Resistivity measurements at hydrostatic pressures to 5.3 GPa were performed in a miniature clamped toroidal cell. A diamond sample and lead manometer were placed inside a small Teflon capsule filled with liquid. Pressure was applied at room temperature and then the clamped cell was cooled slowly to 1.1 K. Measurements of the resistivity in the range 0.5–4 K and in magnetic fields to 4 T were performed in a Quantum Design Physical Properties Measurement System. Measurements of a.c. magnetic susceptibility were made in two apparatuses, one to $^4$He and the other to $^3$He temperatures, by measuring the mutual inductance of a miniature coil with an LR700 bridge. The coil response was calibrated by measuring the susceptibility of a piece of superconducting indium, having a shape and size similar to that of diamond samples, or by normalizing the data to d.c. susceptibility results obtained for $T \geq 1.8$ K in a Quantum Design Magnetic Properties Measurement System with an applied magnetic field of 5 Oe.

**Acknowledgements** We thank D. Wayne for mass spectrometry measurements of the B content of our samples, A. Presz for SEM images and S. Gierlotka for help in sample analysis. This work was supported by the Russian Foundation for Basic Research and by the Strongly Correlated Electrons Program of the Department of Physical Sciences, Russian Academy of Sciences. Work at Los Alamos was performed under the auspices of the US DOE.

**References.**

[1]. Field, J. E. *The Properties of Natural and Synthetic Diamond* (Academic, New York, 1992).

[2]. May, P. W. Diamond thin films: a 21st-century material. *Phil. Trans. R. Soc. Lond. A* **358,** 473–495 (2000).

[3]. Kalish, R. The search for donors in diamond. *Diamond Relat. Mater.* **10,** 1749–1755 (2001).

[4]. Geis, M. W., Twitchell, J. C., Macauley, J. & Okano, K. Electron field emission from diamond and other carbon materials after H2, O2 and Cs treatment. *Appl. Phys. Lett.* **67,** 1328–1330 (1995).

[5]. Teukam, Z. *et al.* Shallow donors with high n-type electrical conductivity in homoepitaxial deuterated boron-doped diamond layers. *Nature Mater.* **2,** 482–486 (2003).

[6]. Garrido, J. A. *et al.* Fabrication of in-plane gate transistors on hydrogenated diamond surfaces. *Appl. Phys. Lett.* **82,** 988–990 (2003).

[7]. Williams, A. W. S., Lightowlers, E. C. & Collins, A. T. Impurity conduction in synthetic semiconducting diamond. *J. Phys. C* **3,** 1727–1735 (1970).

[8]. Vishnevskii, A. S., Gontar', A. G., Torishnii, V. I. & Shul'zhenko, A. A. Electrical conductivity of heavily doped p-type diamond. *Sov. Phys. Semicond.* **15,** 659–661 (1981).

[9]. Werner, M. *et al.* Charge transport in heavily B-doped, polycrystalline diamond films. *Appl. Phys. Lett.* **64,** 595–597 (1994).

[10]. Borst, T. H. & Weis, O. Boron-doped homoepitaxial diamond layers: fabrication, characterization and electronic applications. *Phys. Status Solidi A* **154,** 423–444 (1996).

[11]. Zhang, R. J., Lee, S. T. & Lam, Y. W. Characterization of heavily boron-doped diamond films. *Diamond Relat. Mater.* **5,** 1288–1294 (1996).

[12]. Eremets, M. I., Struzhkin, V. V., Mao, H.-K. & Hemley, R. J. Superconductivity in boron. *Science* **293,** 272–274 (2001).

[13]. Werthamer, N. R., Helfand, E. & Hohenberg, P. C. Temperature and purity dependence of the superconducting critical field $H_{c2}$. III. Electron spin and spin-orbit effects. *Phys. Rev.* **147,** 295–302 (1966).

[14]. McMillan, W. L. Transition temperature of strongly coupled superconductors. *Phys. Rev.* **167,** 331–344 (1968).

[15]. Gurevich, V. L., Larkin, A. I. & Firsov, Yu. A. On the possibility of superconductivity in semiconductors. *Sov. Phys. Solid State* **4,** 131–135 (1962).

[16]. Cohen, M. L. Superconductivity in many-valley semiconductors and in semimetals. *Phys. Rev.* **134,** A511–A521 (1964).

[17]. Cohen, M. L. in *Superconductivity* Vol. 1 (ed. Parks, R. D.) 615–664 (Marcel Dekker, New York, 1969).

[18]. Kohmoto, M. & Takada, Y. Superconductivity from an insulator. *J. Phys. Soc. Jpn* **59,** 1541–1544 (1990).

[19]. Nozières, P. & Pistolesi, F. From semiconductors to superconductors: a simple model for pseudogaps. *Eur. Phys. J. B* **10,** 649–662 (1999).

[20]. Hulm, J. K., Ashkin, M., Deis, D. W. & Jones, C. K. in *Progress in Low Temperature Physics* Vol. VI (ed. Gorter, C. J.) 205–242 (North-Holland, Amsterdam, 1970).

[21]. Kawaji, H., Horie, H., Yamanaka, S. & Ishikawa, M. Superconductivity in the silicon clathrate compound $(Na,Ba)_xSi_{46}$. *Phys. Rev. Lett.* **74,** 1427–1429 (1995).

[22]. Grosche, F. M. *et al.* Superconductivity in the filled cage compounds $Ba_6Ge_{25}$ and $Ba_4Na_2Ge_{25}$. *Phys. Rev. Lett.* **87,** 247003 (2001).

[23]. Khvostantsev, L. G., Vereshchagin, L. F. & Novikov, A. P. Device of toroid type for high pressure generation. *High Temp. High Press.* **9,** 637–639 (1977).

[24]. Voronov, O. A. & Rakhmanina, A. V. Parameter of the cubic cell of diamond doped with boron. *Inorg. Mater.* **29,** 707–710 (1993).

[25]. Brunet, F., Deneuville, A., Germi, P., Pernet, M. & Gheeraert, E. Variation of the cell parameter of polycrystalline boron doped diamond films. *J. Appl. Phys.* **81,** 1120–1125 (1997).

[26]. Bean, C. P. Magnetization of hard superconductors. *Phys. Rev. Lett.* **8,** 250–253 (1962).

**Figure legends**

**Figure 1** Optical and scanning electron microscopy images of the material. Top, central part of the high-pressure synthesis cell after subjecting graphite and $B_4C$ to high-pressure, high-temperature conditions. D, diamond; G, graphite. Bottom, SEM image of B-doped diamond synthesized at high pressures and temperatures.

**Figure 2** X-ray diffraction and Raman spectra of B-doped diamond. **a**, X-ray diffraction pattern of B-doped diamond. The strongest peaks of $B_4C$ at 16–17° and graphite near 12–13° are absent in this pattern from diamond, but are visible in spectra (not shown) of material taken from the boundary between $B_4C$ and diamond. We estimate that unreacted $B_4C$ in the polycrystalline diamond does not exceed 1–2%. The diffraction pattern gives a cubic lattice parameter of 3.5755±0.0005 Å, which is larger than 3.5664 Å for undoped diamond and within the range of lattice parameters 3.575–3.5767 Å for maximally B-doped diamond[24,25]. **b**, Raman spectra of diamond, synthesized in the system graphite–$B_4C$ (lower curve) and CVD films[11] with boron concentrations of $3.4\times10^{17}$ cm$^{-3}$ (upper curve) and $1.5\times10^{21}$ cm$^{-3}$ (middle curve).

**Figure 3** Electrical resistivity and upper critical field curves for B-doped diamond. **a**, Temperature dependence of the electrical resistivity of B-doped diamond at normal and representative high pressures. The insets show details of the resistivity behaviour below 5 K and the pressure-induced shift $\Delta T_c$ of the midpoint of the resistive transition. **b**, Temperature dependence of the upper critical field for B-doped diamond. The resistive mid-point is used to define $H_{c2}$. The inset shows the evolution of the resistivity near $T_c$ at different magnetic fields 0–4 T.

**Figure 4** Magnetic susceptibility of B-doped diamond near $T_c$. The a.c. data (triangles) were normalized to the d.c. data (circles) at 1.8 K. ZFC, zero-field cooled; FC, field cooled. No diamagnetic signal was found in $B_4C$ samples prepared at the same pressure and temperature conditions used for the diamond synthesis. The inset plots magnetization versus field at 1.8 K. From the magnitude of the hysteresis at $H=0$, a simple estimate based on Bean's critical state model[26] gives a critical current density of $\sim 1.45\times10^6$ A m$^{-2}$.

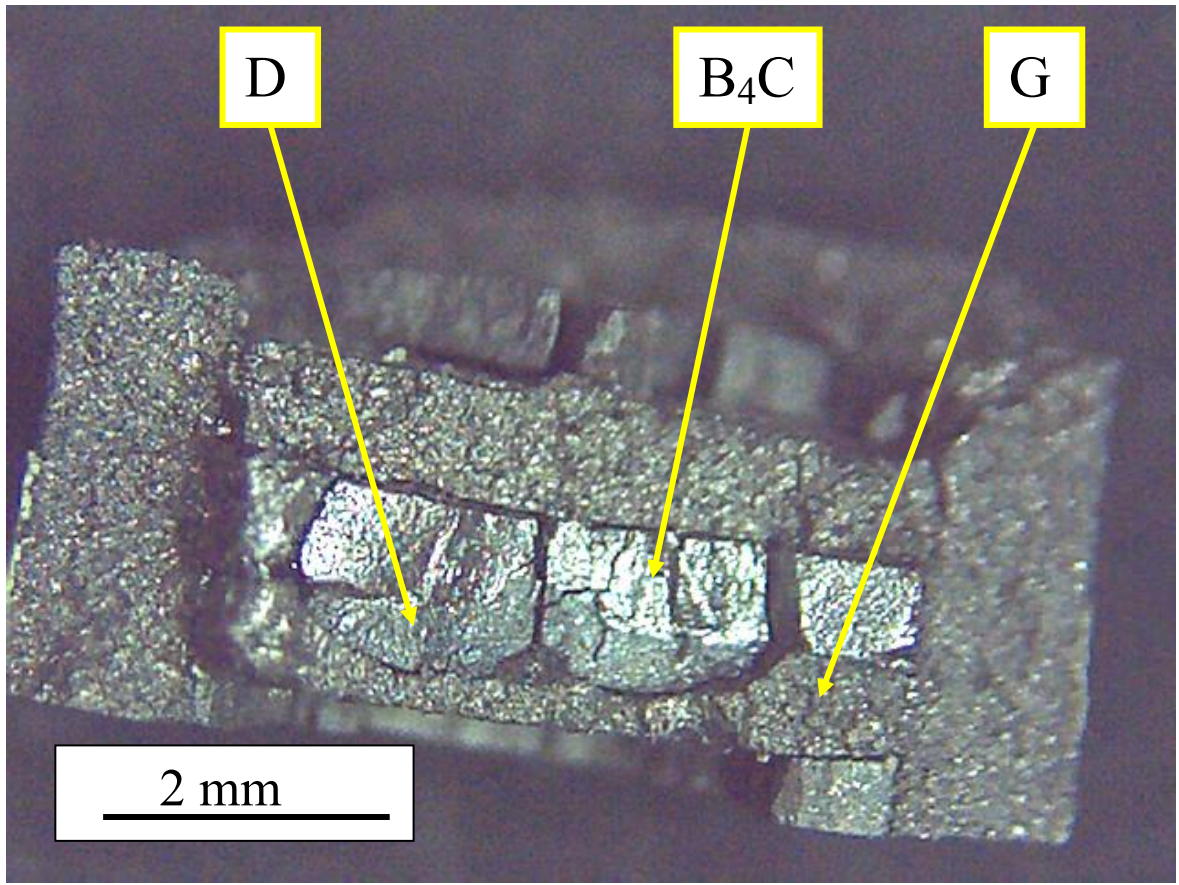

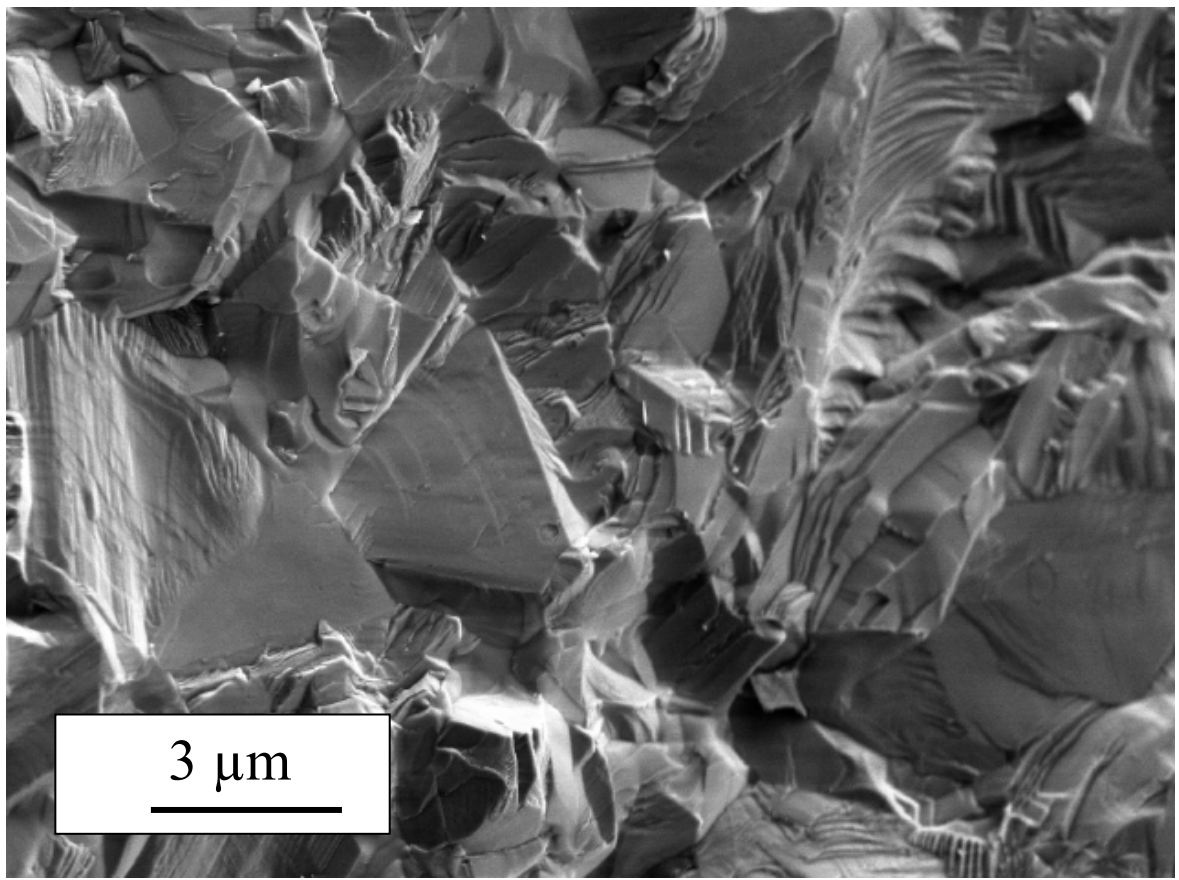

Figure 1

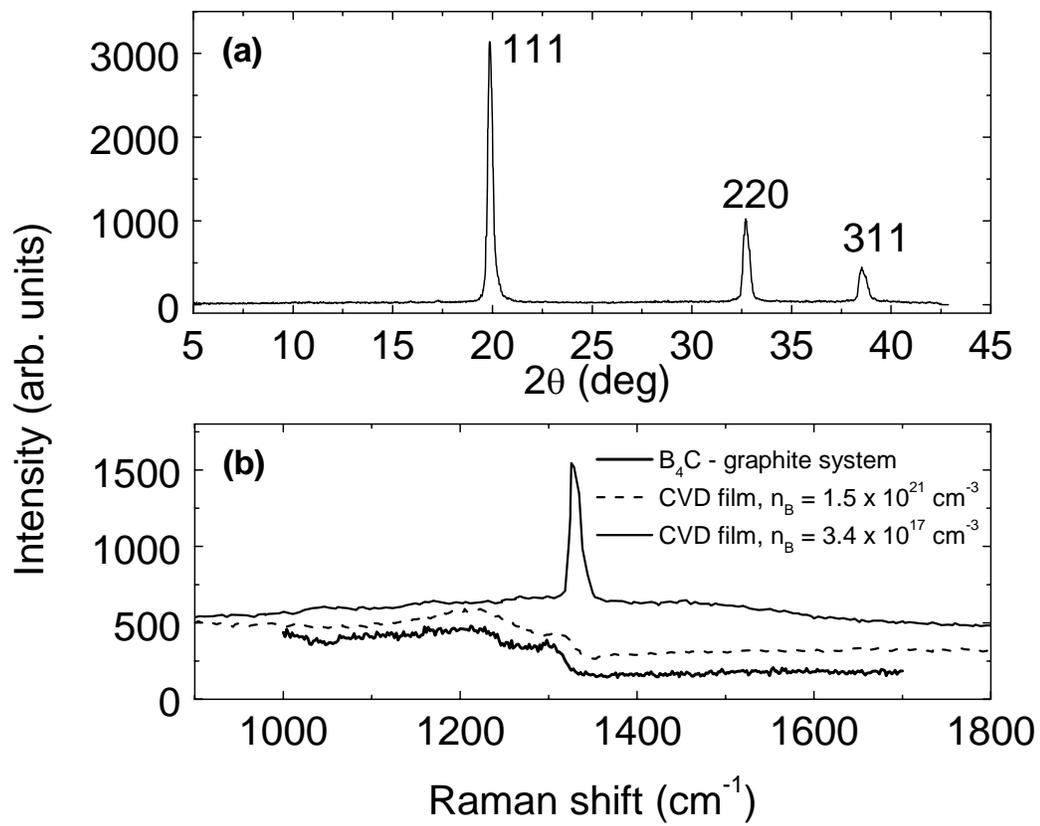

Figure 2

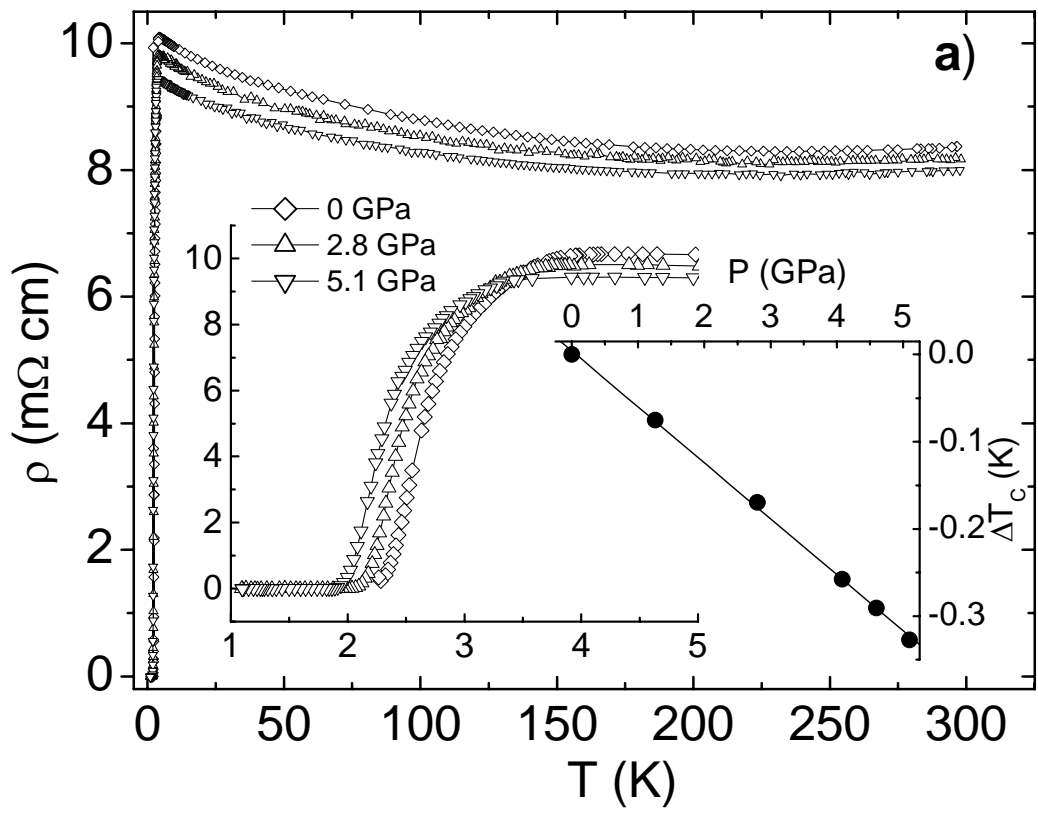

Figure 3a

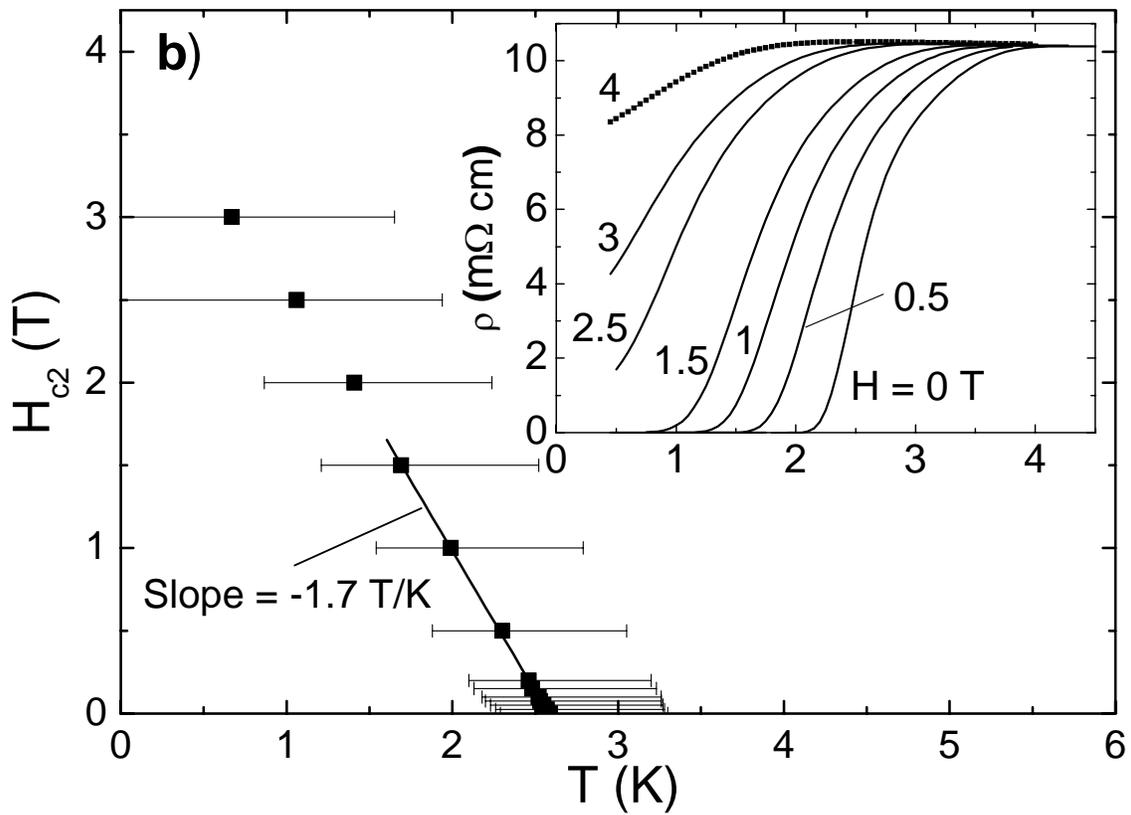

Figure 3b

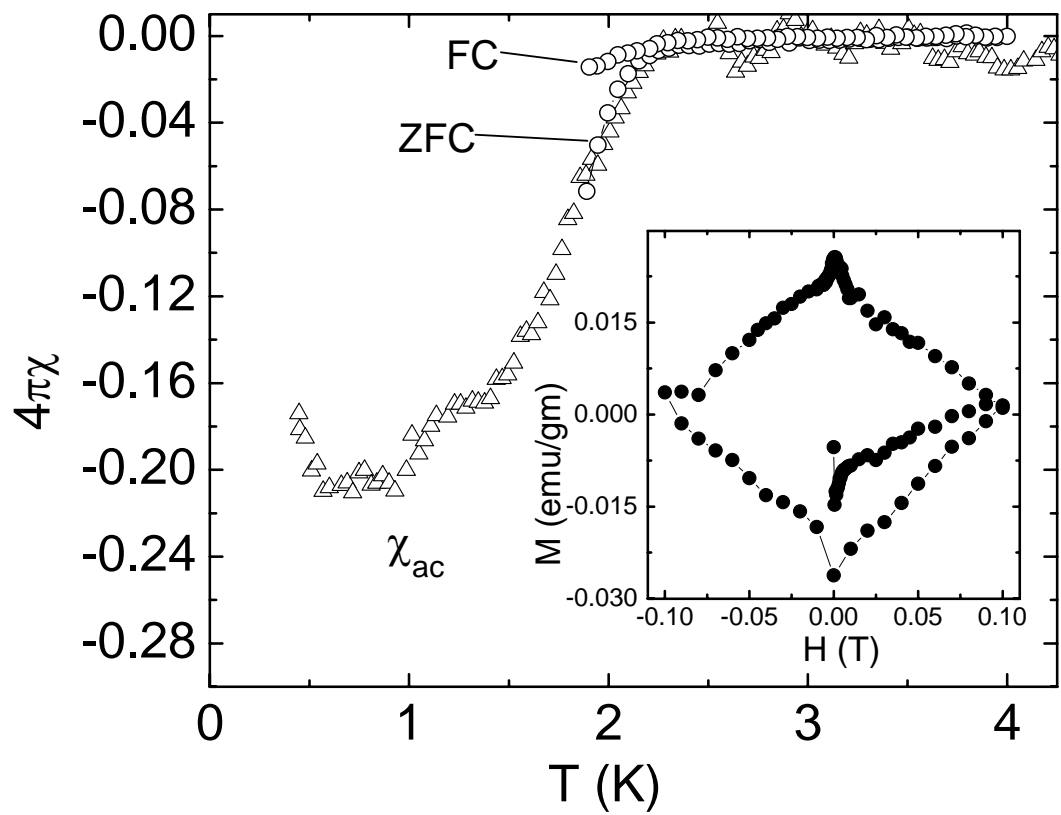

Figure 4